\documentclass[aps,pra,onecolumn,showpacs,superscriptaddress,floatfix]{revtex4}
\usepackage{graphicx,times,nicefrac,amsmath,amsfonts,amssymb,amsthm,epsf,bm,bbm,longtable,dcolumn}
\usepackage{revsymb,wasysym,mathrsfs,color}

\begin{document}

\title{About universal scalings in double K-shell photoionization}

\author{A I Mikhailov}
\affiliation{Petersburg Nuclear Physics Institute, 188300 Gatchina,
St.~Petersburg, Russia}

\affiliation{ Max-Planck-Institut f\"ur Physik komplexer Systeme,
D-01187 Dresden, Germany}

\author{A V Nefiodov}
\affiliation{Petersburg Nuclear Physics Institute, 188300 Gatchina,
St.~Petersburg, Russia}
\affiliation{Institut f\"ur Theoretische
Physik, Technische Universit\"at Dresden, D-01062  Dresden, Germany}

\author{G Plunien}
\affiliation{Institut f\"ur Theoretische Physik, Technische
Universit\"at Dresden, D-01062  Dresden, Germany}

\date{Received \today}
\widetext
\begin{abstract}
We discuss the problem of the universal scalings in the double
ionization of atomic K-shell electrons caused by absorption of a
single photon. In particular, we envisage the following questions:
Under which conditions and up to which accuracy the universal
scalings are realized? Does it make sense to talk about different
physical mechanisms in the double-ionization process? Finally, we
present also the theoretical analysis of recent experimental
measurements performed on neutral atoms. As a testing ground, QED
perturbation theory is employed.
\end{abstract}
\maketitle

The double K-shell photoionization (or the so-called double K-shell
photoeffect) is the fundamental process, which is being persistently
investigated already for more than three decades \cite{1,2}. The
extended interest in the problem is due to the fact that the
two-electron ejection is exclusively caused by the correlation
interactions. Accordingly, it serves as a testing ground for
different theoretical approaches. The most of previous experimental
measurements concerns with the energy behavior for the ratio of
double-to-single photoionization cross sections in neutral helium
\cite{3,4,5,6}. The other atomic targets are investigated much less
thoroughly. In view of the recent developments of novel synchrotron
radiation sources, such experiments become feasible to perform in a
wide range of photon energies \cite{7,8,9,10,11}. Therefore, the
theoretical study of the universal scalings is of particular
importance, because it provides information about most generic
features of the double-ionization process for a whole family of
atomic systems.

\begin{figure}[h]
\centerline{\includegraphics[scale=0.6]{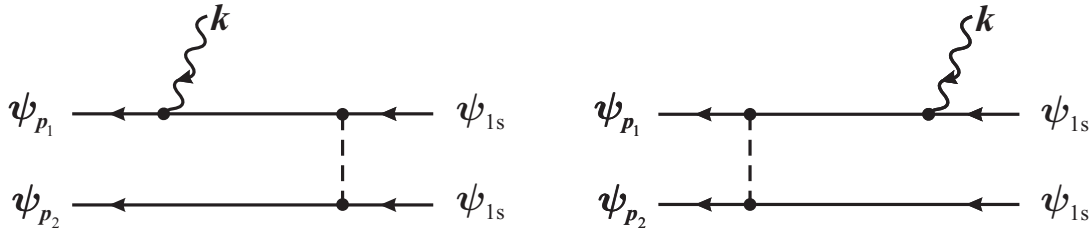}}
\caption{\label{fig1} Feynman diagrams for ionization of the bound
K-shell electrons following by the absorption of a single photon.
Solid lines denote electrons in the Coulomb field of the nucleus,
dashed line denotes the electron-electron Coulomb interaction, and
the wavy line denotes an incident photon.}
\end{figure}

The problem of the double K-shell photoionization of helium-like
ions allows for rigorous theoretical treatment within the framework
of QED perturbation theory \cite{12,13,14,15,16,17,18}.  For
moderate values of the nuclear charge $Z$, the bound K-shell
electrons are characterized by the binding energy $I=\eta^{2}/(2m)$
and the average momentum $\eta=m\alpha Z$, where $\alpha$ is the
fine-structure constant and $m$ is the electron mass ($\hbar=1$,
$c=1$). The incident photon is characterized by the momentum
$\bm{k}$ and the energy $\omega =k$. In the non-relativistic case,
it is assumed that $\alpha Z \ll 1$ and $\omega \ll m$. Since the
atomic nucleus is much heavier than the electron, the former can be
represented as a source of the external field (the Furry picture).
Accordingly, to zeroth approximation, the electrons are described by
the Coulomb wave functions for the discrete and continuous spectra.
The electron-electron interaction is mediated by the exchange of a
different number of virtual photons. Ma\-the\-ma\-ti\-cal\-ly, it is
equivalent to the expansion with respect to the correlation
parameter $Z^{-1}$, which exhibits fast convergence for any
$Z\geqslant 1$ \cite{19}. To leading order, the amplitude of the
double K-shell photoionization consists of two Feynman diagrams,
which describe the electron-electron interactions in the initial and
final states by the one-photon exchange (see Fig.~\ref{fig1}).

It is convenient to distinguish between the near-threshold energy
domain ($\omega \gtrsim I_{2K}$, where $I_{2K}$ denotes the
double-ionization potential for the K shell) and the asymptotic
high-energy one ($I_{2K} \ll \omega \ll m$). In the near-threshold
regime, the theoretical description of the double photoionization is
the most complicated task, because all bindings in the atomic system
are significant. Accordingly, both the electron-electron and
electron-nucleus interactions should be taken into account. The main
contribution to the double-ionization cross section arises, when the
asymptotic momenta $\bm{p}_{1,2}$ of the ejected electrons and the
recoil momentum $\bm{q}$ transferred to the nucleus turn out to be
of the same order of magnitude, namely, $p_1 \sim p_2 \sim q \sim
\eta$, while the photon momentum $k$ is small ($k \ll \eta$). It
implies that the process occurs at atomic distances of the order of
the K-shell radius. In order to study the universal scalings, it is
useful to introduce the dimensionless quantities such as
$\varepsilon_\gamma = \omega/I$ and $\varkappa = k/\eta$. To leading
order of the perturbation theory, $I_{2K}=2I$, that is, the double
K-shell ionization can occur for the photon energy
$\varepsilon_\gamma \geqslant 2$. The dimensionless photon momentum
$\varkappa = \alpha Z \varepsilon_\gamma/2$ serves as a quantitative
measure of the non-dipolarity \cite{15}. If $k \ll \eta$, the dipole
approximation ($\varkappa=0$) can be employed. Obviously, this holds
true within the near-threshold energy domain, if $\alpha Z \ll 1$.

In the asymptotic high-energy regime, some of the electron-nucleus
bindings in the atomic system become to be insignificant, since
another small parameter comes into play, namely,
$\varepsilon_\gamma^{-1}\ll 1$. In this case, the main contribution
to the double-ionization cross section arises in the following
kinematics: $p_1 \sim q \sim \sqrt{2 m \omega} \gg \omega =k$ and
$p_2 \sim \eta$ (or due to the identity of electrons, $p_2 \sim q
\sim \sqrt{2 m \omega} \gg k$ and $p_1 \sim \eta$). In other words,
one ejected electron is fast, acquiring almost all the incident
photon energy. Another ejected electron turns out to be slow. The
photon momentum $k$ is now small compared to the asymptotic momentum
of the fast outgoing electron and, therefore, can be neglected.
Accordingly, the dipole approximation is again legitimate. Due to
the gauge invariance of QED, one can choose the appropriate gauge,
which simplifies the numerical calculations. For example, the
diagram describing the electron-electron interaction in the final
state turns out to be strongly suppressed, provided the Coulomb
gauge is used \cite{12}.

The Feynman diagrams depicted in Fig.~\ref{fig1} represent a
gauge-invariant set, the total contribution of which does not depend
on the gauge of the electron-photon interaction. The separate
contributions of each diagram, which sometimes are referred to as
dominant ``physical mechanisms", are gauge dependent and, therefore,
are devoid of the physical meaning. In many-body perturbation
theory, this has been demonstrated numerically for the particular
case of the helium atom \cite{20,21}. In differential and total
cross sections, which are observable quantities, the diagrams
contribute always coherently, giving rise to the interference terms.
The interference is sometimes neglected in the theoretical analysis,
as if it would be of minor importance (see, for example, recent
papers \cite{7,8,9,10}). However, within the near-threshold energy
domain, this assumption is definitely not correct, because all
correlation contributions are of the same order of magnitude.
Moreover, beyond the leading order, the electron-electron
interactions in the initial and final states turn out to be
generally entangled due to the crossed photon diagrams (the vertex
correction), although the higher-order correlation contributions are
suppressed by the parameter $Z^{-1}$. As a consequence, the attempts
to separate the ``physical mechanisms" fail. The discussion of the
double photoionization in terms of ``physical mechanisms" is
superfluous, because the concepts themselves are physically
unfounded.

\begin{figure}[h]
\begin{minipage}[t]{0.49\textwidth}
\centering\includegraphics[width=0.9\textwidth,angle=0,clip]{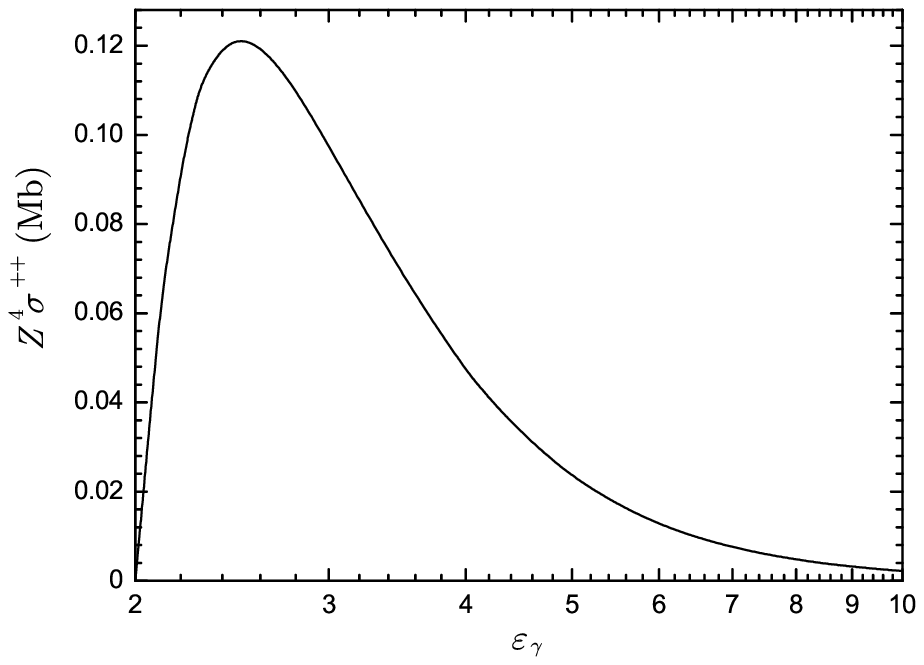}
\end{minipage}
\begin{minipage}[t]{0.49\textwidth}
\centering\includegraphics[width=0.9\textwidth,angle=0,clip]{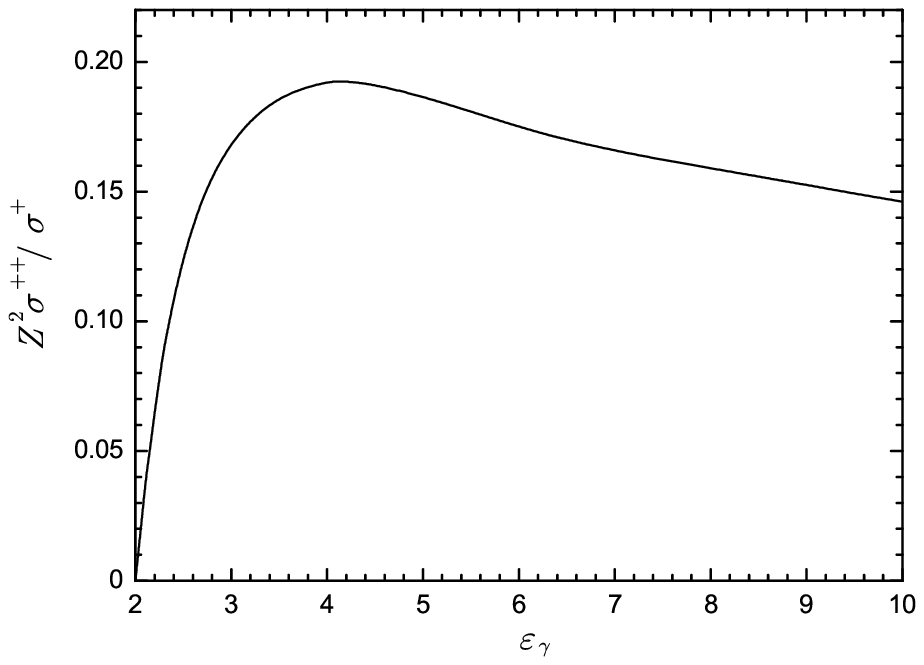}
\end{minipage}
\caption{\label{fig2} The universal functions $Z^{4}\sigma^{++}$ and
$Z^{2}\sigma^{++}/\sigma^{+}$ calculated with respect to the
dimensionless photon energy $\varepsilon_\gamma$ \cite{16,17}.}
\end{figure}

To leading orders of QED perturbation theory with respect to the
parameters $\alpha Z$ and $Z^{-1}$, the total cross sections for
double and single K-shell photoionizations are given by \cite{16,17}
\begin{eqnarray}
\sigma^{++} &=& \sigma_{0}Q(\varepsilon_\gamma,\varkappa)Z^{-4} \, ,
\label{eq1}\\
\sigma^{+}  &=&  \sigma_{0}F(\varepsilon_\gamma,\varkappa)Z^{-2}\,
, \label{eq2}
\end{eqnarray}
where $\sigma_{0}= \alpha \pi a_0^2$ and $a_0=1/(m \alpha)$ is the
Bohr radius. The dimensionless functions
$Q(\varepsilon_\gamma,\varkappa)$ and
$F(\varepsilon_\gamma,\varkappa)$ depend on the nuclear charge $Z$
via the incident photon momentum $\varkappa$. The scalings
\eqref{eq1} and \eqref{eq2} become to be universal, if one sets
$\varkappa=0$ only. Accordingly, the quantities $\alpha Z$,
$Z^{-1}$, and $\varkappa$ estimate the accuracy, with which the
universal scalings take place. In the dipole approximation, the
function $F(\varepsilon_\gamma)$ is known in the analytical form
\cite{22}
\begin{equation}\label{eq3}
F(\varepsilon_\gamma)= \frac{2^{10} \pi}{3
\varepsilon_\gamma^{4}}\frac{\exp(-4 \xi \cot^{-1}\xi)}{[1- \exp (-2
\pi \xi)]} \,  ,
\end{equation}
where $\xi = 1/\sqrt{\varepsilon_\gamma -1}$. The universal function
$Q(\varepsilon_\gamma)$ can be obtained by the numerical integration
only (see Fig.~\ref{fig2}). The calculations performed by
independent theoretical groups are consistent with each other
\cite{16,17,18}. Note also that the maxima of the universal
functions $Z^{4}\sigma^{++}$ and $Z^{2}\sigma^{++}/\sigma^{+}$ peak
at different values of the photon energy, namely, at
$\varepsilon_\gamma \simeq 2.5$ and 4, respectively. In the
asymptotic non-relativistic range characterized by the condition $2
\ll \varepsilon_\gamma \ll 2 (\alpha Z)^{-2}$, the double-to-single
photoionization ratio does not depend on the photon energy
\cite{12}. To a given order of the non-relativistic perturbation
theory with respect to the electron-electron interaction, it is
represented as the Pad\'{e} approximant \cite{19}. The ratio
$\sigma^{++}/\sigma^{+}$ can be less sensitive to the correlation
effects rather than the cross sections themselves, when the
higher-order corrections to $\sigma^{++}$ and $\sigma^{+}$ have the
same signs. In the relativistic limit, which is characterized by
$\varepsilon_\gamma \gtrsim 2 (\alpha Z)^{-2}$, the double-to-single
photoionization ratio again depends on the incident photon energy
\cite{13,14}.

In the case of neutral atoms, the universal scaling behavior deduced
for helium-like targets still remains to be valid, if one takes into
account the screening effect of the outer-shell electrons \cite{16}.
This can be achieved by substitution of the true nuclear charge $Z$
by its effective value $Z_{\textrm{eff}}$, which is defined via the
experimentally observable (or Hartree-Fock) ionization potential
$I_{\textrm{exp}}=m(\alpha Z_{\textrm{eff}})^2/2$. In addition, the
photon energy $\omega$ should be calibrated in units of
$I_{\textrm{exp}}$, namely, $\varepsilon_\gamma = \omega /
I_{\textrm{exp}}$. This procedure was also adopted by Hoszowska J
{\it et al} \cite{10,11}, although not properly cited.

In works \cite{16,17,18}, the theoretical analysis is made for the
experimental data on the double K-shell photoionization available
until 2004. A satisfactory agreement of the universal scaling and
experimental results is found in the majority of cases. However, for
neutral molybdenum ($Z=42$), the measurements performed at $\omega =
50$ keV give the double-to-single photoionization ratio $\sim 3.4(6)
\times 10^{-4}$ \cite{23}, which is more than three times as large
as the theoretical prediction $\sim 0.87 \times 10^{-4}$
\cite{16,17,18}. Recently, a similar disagreement between the theory
and experimental data has been also found for silver atoms ($Z=47$)
\cite{7}. Of course, for such heavy targets, the corrections with
respect to parameters $\alpha Z$ and $\varkappa$ can become
significant. However, the disagreement of the universal scaling with
the experimental results of papers~\cite{7,23} cannot be explained
by the account for higher-order corrections, because the
disagreement is much larger than the parametrical values $\alpha Z$,
$Z^{-1}$, and $\varkappa$. In addition, as we argue below, the
results of works~\cite{7,23} are in contradiction not only with the
theory, but also with other experimental measurements, for example,
with work~\cite{24}, where the double K-shell photoionization is
measured for Cu atoms. Note that the results of Diamant R {\it et
al}~\cite{24} are consistent with the universal curve
\cite{16,17,18}. Therefore, this remains as a challenge for further
experimental and theoretical investigations.

The obvious consequence of the universal scalings consists in the
following. For neutral atoms, at any fixed value of the
dimensionless photon energy $\varepsilon_\gamma$, the
double-to-single photoionization ratio $\sigma^{++}/\sigma^{+}$
should decrease, if the effective nuclear charge $Z_{\textrm{eff}}$
(and, respectively, the value of $Z$) increases. Indeed, at any
particular value $\varepsilon_\gamma$, the ratio is given by
$\sigma^{++}/\sigma^{+}=C Z^{-2}_{\textrm{eff}}$, where $C$ is a
constant. However, if we compare, for example, the experimental
ratio $\sigma^{++}/\sigma^{+}$ for Mo \cite{23} and Cu \cite{24}
near the maxima of the double K-shell photoionization cross
sections, it turns out to be significantly larger for heavier Mo
rather than that for Cu. A similar conclusion can also be drawn for
measurements performed on Ag atoms \cite{7}. Accordingly, the
experimental data \cite{7,23} seem to be not quite reliable. One of
the possible reasons, which could perhaps explain the strong
discrepancy between theory and experiment, might be the contribution
of some other (indirect) ionization channels.

\begin{figure}[h]
\begin{minipage}[t]{0.49\textwidth}
\centering\includegraphics[width=0.9\textwidth,angle=0,clip]{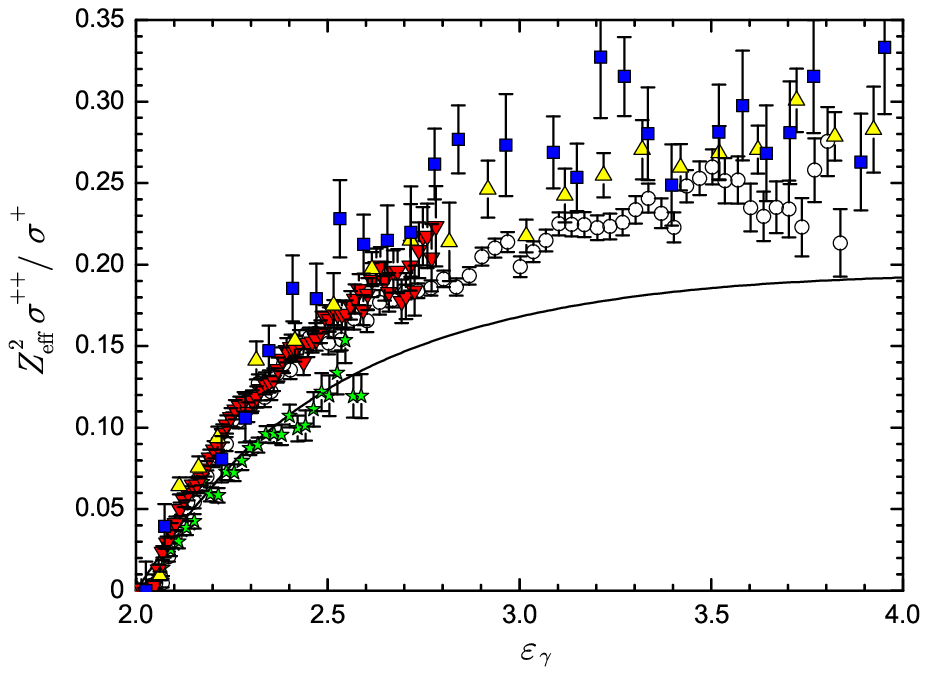}
\end{minipage}
\begin{minipage}[t]{0.49\textwidth}
\centering\includegraphics[width=0.9\textwidth,angle=0,clip]{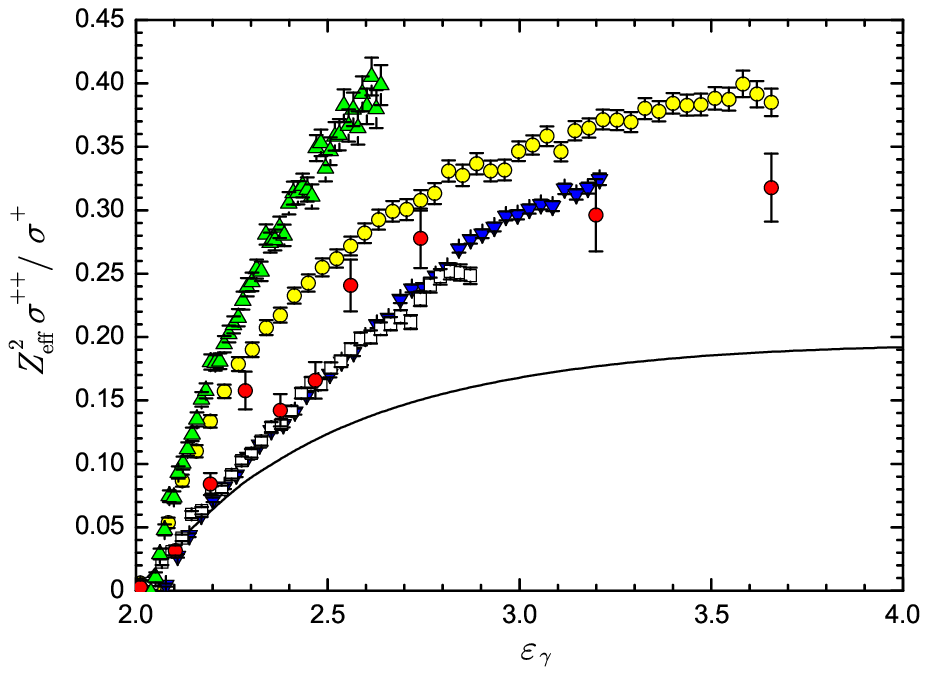}
\end{minipage}
\caption{\label{fig3} (Color online) The universal ratio of
double-to-single K-shell photoionization cross sections
\cite{16,17}. Measurements by Huotari S {\it et al} \cite{8}: V,
\textcolor{yellow}{$\newmoon$}; Cr, $\fullmoon$; Mn,
\textcolor{blue}{$\blacktriangledown$}; Co, $\square$; Ni,
\textcolor{green}{$\blacktriangle$}; Cu,
\textcolor{red}{$\blacktriangledown$}; Zn,
\textcolor{green}{$\bigstar$}. Measurements by Oura M {\it et al}
\cite{25}: Ca, \textcolor{blue}{$\blacksquare$}; Ti,
\textcolor{yellow}{$\blacktriangle$}; V,
\textcolor{red}{$\newmoon$}.}
\end{figure}

In paper \cite{8}, the near-threshold energy behavior for the
double-to-single K-shell photoionization cross section ratios is
studied for neutral atoms with nuclear charges within the range
$23\leqslant Z \leqslant 30$. These experimental results, however,
rise more new questions, rather than giving affirmative answers. One
part of the results is in fair agreement with the theoretical
universal scaling, while another part of the measurements is in
disagreement with it (see Fig.~\ref{fig3}). Since the disagreements
are quite significant, they cannot be explained by account for
higher-order corrections with respect to the small parameters
$\alpha Z$, $Z^{-1}$, and $\varkappa$. The slope of the experimental
curves for the neighboring elements changes drastically at the
transition from Ni to Cu, but undergoes rather slight changes at the
next transition from Cu to Zn. The physical reasons for such
different behavior of the curves are not clear. A comparison of the
measurements at $\varepsilon_\gamma \simeq 2.5$ reveals that the
ratio $\sigma^{++}/\sigma^{+}$ for Ni atoms is larger than those for
lower-$Z$ atoms. As we have argued above, this indicates on
contradiction of the measurements themselves. Since all experimental
data for V, Mn, Co, and Ni atoms overestimate the theoretical
universal curve for the direct process, the disagreements might be
due to additional contributions of indirect ionization channels. The
experimental data for Zn, the heaviest element studied in work
\cite{8}, exhibit a satisfactory agreement with the non-relativistic
calculations. This indicates that the relativistic effects are of
minor importance. In Fig.~\ref{fig3}, we have also presented the
experimental data for neutral Ca, Ti, and V reported by Oura M {\it
et al} \cite{25}. For V atoms, the measurements performed by
different groups overlap only partly. The quoted error bars seem to
be too small.

\begin{figure}[h]
\centerline{\includegraphics[scale=0.9]{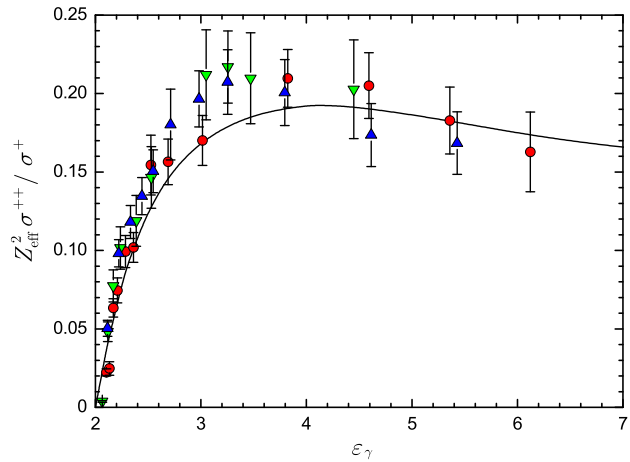}}
\caption{\label{fig4} (Color online) The universal ratio of
double-to-single K-shell photoionization cross sections
\cite{16,17}. Experimental data: Mg, \textcolor{red}{$\newmoon$};
Al, \textcolor{green}{$\blacktriangledown$}; Si,
\textcolor{blue}{$\blacktriangle$} \cite{9,10}.}
\end{figure}

One should stress again that further account for the relativistic,
correlation, and non-dipole corrections will distort the universal
behavior of the ionization cross sections. Therefore, neither the
neutral helium nor very high-$Z$ atoms are the most suitable targets
to test the universality. In the first case, the higher-order
correlation corrections should be taken into account, while the
relativistic and non-dipole corrections are significant in the
second case. The universal scalings have the highest accuracy for
atomic systems with moderate values of the nuclear charge $Z$,
because in this case, all the characteristic parameters $\alpha Z$,
$Z^{-1}$, and $\varkappa$ are small enough. The helium-like
multicharged ions would be the simplest targets, which are clean
from the screening effect of the outer-shell electrons. Up to now,
such measurements are not available in the literature. Recently, the
ratios of double-to-single K-shell photoionization cross sections
are reported for neutral Mg, Al, and Si atoms \cite{9,10}. As seen
from Fig.~\ref{fig4}, these measurements confirm the universal
scaling deduced to leading orders of QED perturbation theory.
However, more experimental information is required in order to
elucidate the generic features of the double ionization of innermost
bound electrons.

\acknowledgements

We thank M Oura and M Deutsch for kind supply of the experimental
data in the numerical form. AM ack\-now\-led\-ges financial support
from Max Planck Institute for the Physics of Complex Systems. AN is
grateful for support to the Alexander von Humboldt Foundation. The
work was partly financed by RFBR under Grant No. 08-02-00460-a.

\end{document}